\documentclass[twocolumn,showpacs,preprintnumbers]{revtex4}

\usepackage{graphicx}
\usepackage{dcolumn}
\usepackage{bm}
\usepackage{amssymb}
\usepackage{amsmath}
\usepackage{epsfig}

\def\cosz {\cos\theta^{\rm zenith}}
\def\lapp{\mathrel{\rlap{\raise.5ex\hbox{$<$}}
                    {\lower.5ex\hbox{$\sim$}}}}
\def\gapp{\mathrel{\rlap{\raise.5ex\hbox{$>$}}
                    {\lower.5ex\hbox{$\sim$}}}}

\def\bc {\begin{center}}
\def\ec {\end{center}}
\def\be {\begin{equation}}
\def\bea {\begin{eqnarray}}
\def\ee {\end{equation}}
\def\eea {\end{eqnarray}}

\begin{document}

\title { 
Prospects of measuring the leptonic CP phase with atmospheric neutrinos
}

\author
{\sf Abhijit Samanta
\footnote{E-mail address: abhijit@hri.res.in}}
\affiliation
{\em Harish-Chandra Research Institute,
Chhatnag Road,
Jhusi,
Allahabad 211 019,
India}


\begin{abstract}
\noindent
We have studied the prospects of measuring the CP violating phase  with atmospheric neutrinos
at a large magnetized iron calorimeter detector considering the muons (directly measurable) 
of the neutrino events generated by a MonteCarlo event generator Nuance. The effect 
of $\theta_{13}$ and $\delta_{CP}$ appears dominantly neither in atmospheric neutrino 
oscillation nor in solar neutrino oscillation, but appears as subleading  in both cases. 
These are observable in range of $E \sim 1$ GeV for atmospheric neutrino, where solar and 
atmospheric oscillation couple. In this regime, the quasi-elastic events dominate and the 
energy resolution is very good, but the angular resolution is very poor. Unlike beam experiments 
this poor angular resolution acts against its measurements. However, we find that one can be able 
to distinguish $\delta_{CP}\approx 0^\circ$ and $180^\circ$  at 90\% confidence level. We find 
no significant sensitivity for $\delta_{CP}\approx 90^\circ$ or $270^\circ$.
\end{abstract}


\maketitle
\section{Introduction}
The evidence of neutrino masses and their mixing through neutrino oscillation \cite{Fukuda:1998mi}
 is the first footprint onto the physics beyond the standard model (SM). However, the first
hint was obtained  observing the anomaly by IMB in 1986 and then confirmed 
by Kamiokande in 1988 \cite{Haines:1986yf, Hirata:1988uy}.
The oscillation experiments provide the measurements of mass squared differences
 and mixing
angles.
At present $1 (3) \sigma$ ranges are \cite{Fogli:2008ig}:
$\Delta m_{21}^2 = 7.67^{+0.16}_{-0.19} (^{+0.42}_{-0.53}),
~~|\Delta m_{31}^2| = 2.39^{+0.11}_{-0.8} (^{+0.42}_{-0.33}),
~~\sin^2\theta_{12}=0.312^{+0.019}_{-0.018} (^{+0.063}_{-0.049}),
~~\sin^2\theta_{23}=0.466^{+0.058}_{-0.073} (^{+0.178}_{-0.135}),~~ {\rm and}
~~\sin^2\theta_{13}=0.016^{+0.010}_{-0.010} (<0.046)$.
Here $\theta_{ij}$ are the mixing angles in Pontecorvo, Maki, Nakagawa, Sakata (PMNS)
mixing matrix \cite{mixing} and
       $\Delta m_{ji}^2=m_j^2-m_i^2$.
Currently, there is no constraint on the
CP violating phase $\delta_{CP}$ or on the sign of $\Delta m_{31}^2$.
The origin of CP violation is still an open problem 
that needs both theoretical and experimental exploration. Till now the best tests  come from only 
neutral kaon oscillations. 

The mixing angle $\theta_{13}$ tells how strongly the atmospheric and solar oscillations 
couple and therefore also determines the CP violation effects in neutrino oscillation.
Both $\theta_{13}$ and $\delta_{CP}$ are observable in solar and atmospheric neutrino
oscillation experiments only as subleading. The three flavor effects are masked mainly 
by wide resolutions and  systematic uncertainties. There are enormous efforts to probe
CP violation and to measure $\theta_{13}$ in long baseline reactor experiments Double Chooz 
\cite{Ardellier:2004ui}, Daya bay \cite{Guo:2007ug}, RENO \cite{Kim:2008zzb} and 
accelerator experiments T2K \cite{Itow:2001ee} and NO$\nu$A \cite{Ayres:2004js}.
In a recent paper \cite{Huber:2009cw}, it is shown that there will be only a hint of CP-violation at 90\% 
confidence level for $\sin^22\theta_{13} \gapp 0.05$ for most values of $\delta_{CP}$ 
if one  considers only the upgraded beams for T2K and NOvA combined with reactor data.




The CP violating phase $\delta_{CP}$ (Dirac) has been studied for a magnetized Iron CALorimeter (ICAL) 
detector with atmospheric neutrinos in \cite{Gandhi:2007td}, but there is no 
significant sensitivity. 
In this paper, we have re-examined the sensitivity of $\delta_{CP}$. Then we point out the 
difference from the past analysis. The main difference is that they considered a high 
threshold (2 GeV) in their analysis. For events beyond this energy, the sensitivity almost vanishes.
In this work, we consider only muons 
(directly measurable) of the events generated by a MonteCarlo event generator Nuance-v3 
\cite{Casper:2002sd} and the muon threshold energy of 0.8 GeV. 
The main advantage  of a magnetized detector is that it can measure the oscillation effect 
separately for muon neutrinos and anti-neutrinos.
This type of  detector has been proposed 
by India-based Neutrino Observatory (INO) \cite{Athar:2006yb}.




\section{Atmospheric neutrinos}\label{s:flux}
The atmospheric neutrinos are produced by the interactions of cosmic rays with the
atmosphere. 
These are primarily protons and Heliums and some heavy ions. They produce mainly 
muons, pions and kaons, which decay into neutrinos.  Here we discuss briefly
the uncertainties which  arise in flux calculation.
For calculation of the neutrino flux, one needs the detailed information on 
(i) the primary cosmic-ray spectra at the top of the atmosphere,
(ii) the hadronic interactions between cosmic rays and atmospheric nuclei, 
(iii) the propagation of cosmic-ray particles inside the atmosphere, and
(iv) the decay of the secondary particles.   
The uncertainty on primary cosmic-ray spectra  has been improved by the experiments \cite{Alcaraz:2000vp, Alcaraz:2000zz,Sanuki:2000wh,Boezio:2002ha}.  The hadronic interactions are measured from accelerator experiments. However,
the available data do not cover all required phase space for calculation of the neutrino flux.
The propagation of cosmic-ray particles inside the atmosphere, and the decay of the secondary 
particles are handled accurately by simulation \cite{Sanuki:2006yd}. 

Then the total uncertainty can be expressed   as \cite{Sanuki:2006yd}
$$\delta^2_{\rm total}=\delta^2_{\pi}+\delta^2_{K}+\delta^2_{\sigma}+\delta^2_{\rm air} + ..., $$
where $\delta_\pi$ is the uncertainty due  to uncertainty of $\pi$ production in the hadronic 
interaction model, $\delta_K$ due to $K$ production, $\delta_{\sigma}$  due to hadronic 
interaction cross section and $\delta_{\rm air}$ due to atmospheric density profile.

For calculating atmospheric neutrino flux from
primary cosmic ray flux, the measured atmospheric muon flux is compared with the calculation \cite{Sanuki:2006yd}.
The error in $\mu$ and $\nu$ fluxes comes mainly in the $\pi$ production in the hadronic 
interaction model. It has been shown  that in case of pions above 1 GeV \cite{Sanuki:2006yd}, 
$$\frac{\Delta \phi_\mu}{\phi_\mu}\simeq \frac{\Delta \phi_{\nu_\mu}}{\phi_{\nu_\mu}}\simeq \frac{\Delta \phi_{\nu_e}}{\phi_{\nu_e}}$$
However, this equation does not hold for $E \lapp 1$ GeV. 
The present  estimated uncertainty of atmospheric neutrino flux is discussed later in section
\ref{s:syst}.

\begin{figure*}[htb]
\hspace*{-1.cm}
\includegraphics[height=19.9cm,width=19.cm,angle=270]{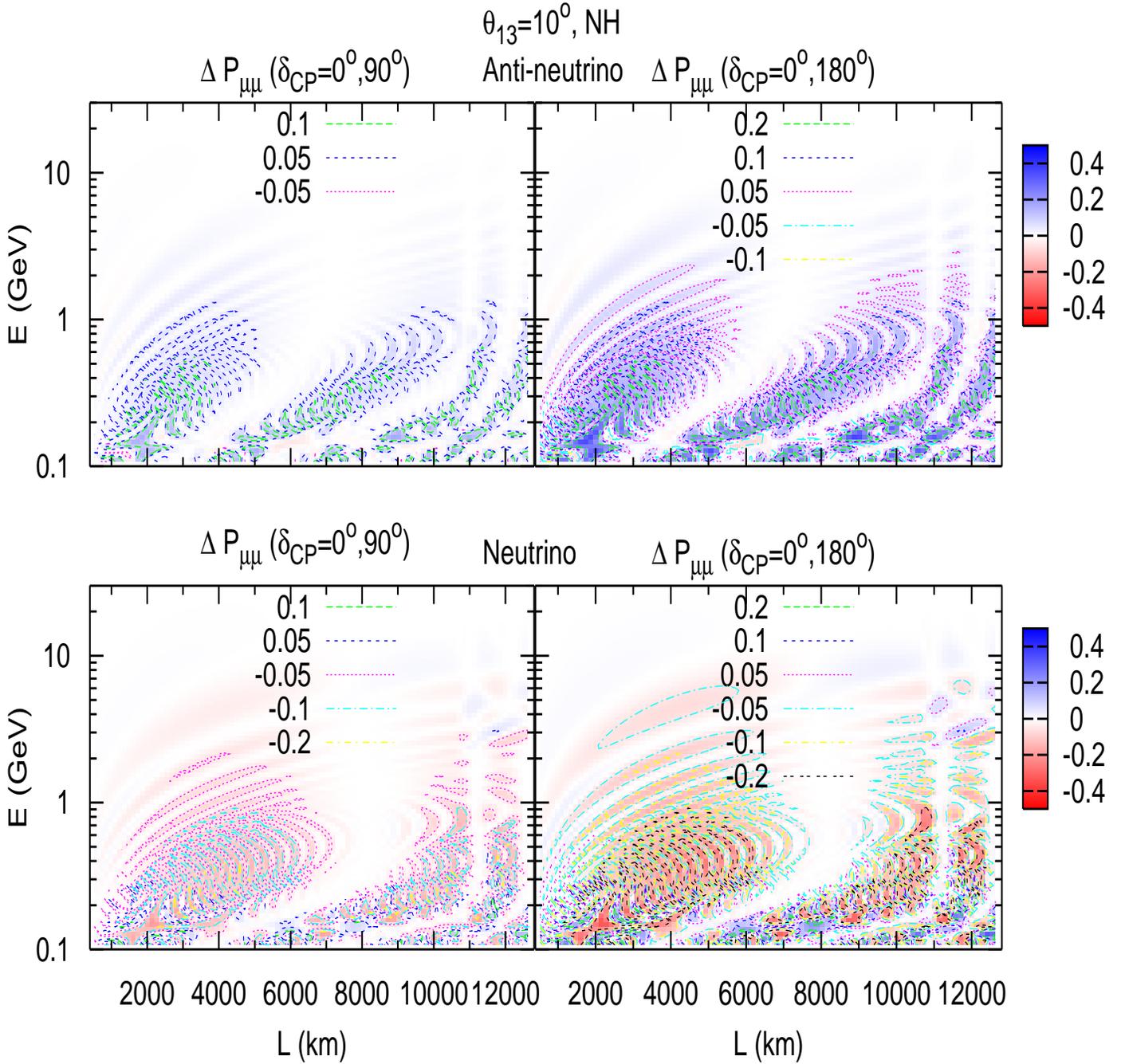}
\caption{\sf \small The oscillogram of the difference $\Delta P
=P_{\mu\mu}(\delta_{CP}=x^\circ) - P_{\mu\mu}(\delta_{CP}=0^\circ)$
in $E-L$ plane for $x=90^\circ$ and $180^\circ$ with $\theta_{13}=10^\circ$ for both
neutrino and anti-neutrino with NH.
We  set other parameter same as fixed for the analysis and given in the text. 
}
\label{f:nh}
\end{figure*}

\begin{figure*}[htb]
\hspace*{-1.cm}
\includegraphics[height=19.cm,width=19.cm,angle=270]{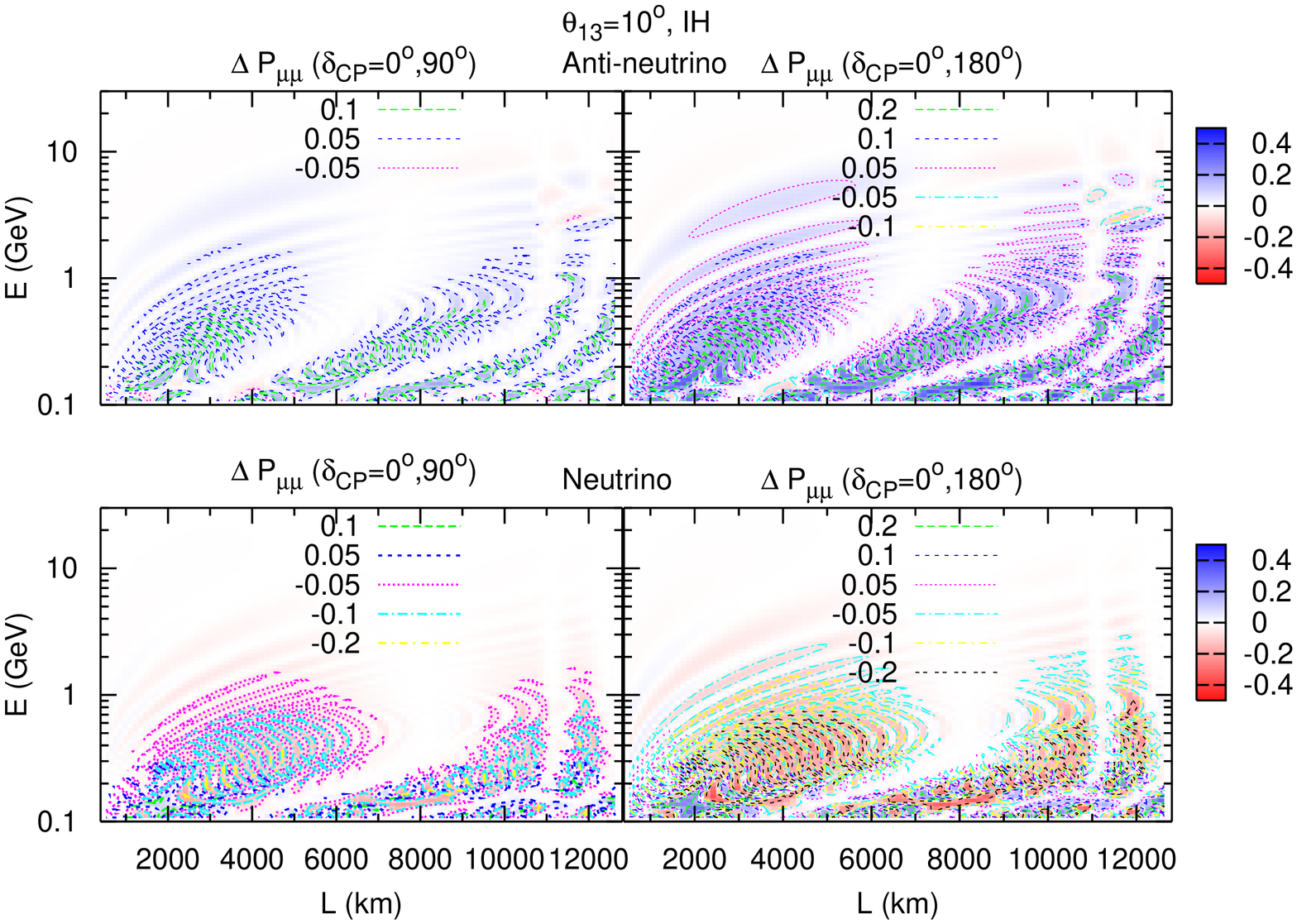}
\caption{\sf \small The same as \ref{f:nh}, but for IH.
}
\label{f:ih}
\end{figure*}

\section{The $\delta_{CP}$ dependence of oscillation probability} 
The algebraic expression for full three flavor oscillation probability with CP phase
$\delta_{CP}$ is very long and complicated to understand its change with the
oscillation parameters. 
We have plotted the oscillogram of 
$\Delta P=[P_{\mu\mu}(\delta_{CP}=x^\circ) - P_{\mu\mu}(\delta_{CP}=0^\circ)]$
in $E-L$ plane with $x=90^\circ$ and $180^\circ$ for $\theta_{13}=10^\circ$. 
We set $|\Delta m_{32}^2|=2.5\times 10^{-3}$eV$^2$, $\theta_{23}=45^\circ$
and the solar parameters at their best-fit values. Since $\delta_{CP}$  appears
together with $\sin\theta_{13}$ in PMNS matrix, the effect  $\delta_{CP}$ in $P_{\mu\mu}$  increases
with increase in  $\sin\theta_{13}$. The solar and atmospheric oscillation are 
coupled through  $\theta_{13}$. The solar neutrino oscillation is dominant
at $E\sim$ a few tens of MeV, while the atmospheric neutrino oscillation 
is dominant at $E\sim$  GeV. These are determined from the corresponding mass squared differences.
For this reason we see that
the $\delta_{CP}$ effect is prominent in sub-GeV region and decreases rapidly
with increase in energy. 
Another interesting feature of the plots is that over a large regions of $E-L$ plane,
$\Delta P$ is positive for anti-neutrino and negative for neutrino in both cases for inverted hierarchy
(IH)  and normal hierarchy (NH).
This is why the poor resolutions cannot wash out fully the CP sensitivity.
It is also expected that the effect of the CP phase is not fully nullified due to the systematic
uncertainties with pull method of chi-square analysis. The reason is following.
The systematic uncertainties are high compared to the difference $\Delta P$.
But, they are constant over a large region of $E-L$ plane, while  $\Delta P$ changes with $L$ and $E$.
However, from these plots one can also find the optimized CP sensitive $E$ and $L$ values
for the experiments with neutrino beams.

\section{The chi-square analysis}
Due to the low statistics at high energy, the $\chi^2$ is calculated
according to the Poisson probability distribution  defined by the
expression:
 \begin{eqnarray}
 \chi^2 &=& \sum_{i,j=1}^{n_L,n_E} \left[ 2 \left\{ N^{p}_{ij} \left(
 1+\sum_{k=1}^{n_s} f^k_{ij} \cdot
 \xi^k \right) - N^{o}_{ij} \right\} \right. \nonumber \\
  && \left. - 2 N^{o}_{ij} \ln \left( \frac{N^{p}_{ij} \left( 1+
 \sum_{k=1}^{n_s} f^k_{ij} \cdot \xi_k \right) } {N^{o}_{ij}}\right)  \right] 
   + \sum_{k=1}^{n_s} {\xi_k}^2 \nonumber
 \label{e:chisq}
 \end{eqnarray}

The $N^o_{ij}$ is considered as the number of observed events generated
by Nuance
for a set of oscillation parameters with an exposure of 1 Mton.year of ICAL.
The $N^p_{ij}$ is the corresponding  number of predicted events (discussed later).
These are obtained in a 2-dimensional grids
in the plane of $\log E - L^{0.4}$.  The $f^k_{ij}$ is the systematic uncertainty
of $N^p_{ij}$ due to the $k$th uncertainty (discussed later).  The ${\xi_k}$ is the pull
variable for the $k$th systematic uncertainty.

The events are binned in the grid of $\log E- L^{0.4}$
plane. We use total number of $\log E$ bins $n_E$ = 35 (0.8 $-$ 40 GeV) and the number of  $L^{0.4}$ bins
$n_L$  as a function of the energy.
We consider $n_L = 2\times 25,~2\times 27,~2\times 29,~2\times 31,$ and $~2\times 33$, when
$E= 0.8-1,~ 1-2,~ 2-3,~ 3-4,~ {\rm and} ~ >4$ GeV, respectively.
For the down-going events, the  binning  is done
by replacing `$L^{0.4}$' by $`- L^{0.4}$'. 
We choose this type of binning to reflect oscillation
effect in a better way in the chi-square analysis \cite{Samanta:2008af,Samanta:2008ag}.
The binning has been optimized in 
\cite{Samanta:2008af,Samanta:2008ag}   for the precision study of $\Delta m^2_{32}$ and
$\theta_{23}$.
We consider the mirror $L$ for down going events, which is same if the neutrino comes from exactly
opposite direction.
The factor `2' is to consider both up and down going cases.
We consider number of events in a bin $> 4$ to maintain $\chi^2/d.o.f \approx 1$ \cite{Samanta:2008af}.
We have checked that this always keeps it $\lapp 1.08$  at the minima in all cases.

\subsection{Systematic uncertainties}\label{s:syst}
We divide the systematic uncertainties into two categories:
I) overall uncertainties (which are independent of
energy and zenith angle), and
II) tilt  uncertainties (which are  dependent of  energy and/or zenith angle).
 The energy dependent flux uncertainty which arises due to the uncertainty in
spectral indices, can be expressed as
\begin{equation}
      \Phi_{\delta_E}(E) = \Phi_0(E) \left( \frac{E}{E_0} \right)^{\delta_E}
      \approx \Phi_0(E) \left[ 1 + \delta_E \log_{10} \frac{E}{E_0} \right] \nonumber\\
\label{e:uncer}
\end{equation}
b)
Similarly, the flux uncertainty as a function of  zenith angle  can be expressed as
\begin{equation}
      \Phi_{\delta_z}(\cos\theta_z)
      \approx \Phi_0(\cos\theta_z) \left[ 1 + \delta_z |\cosz -\cos\theta_0^{\rm zenith}| \right]   \nonumber\\
\end{equation}
c) overall flux normalization uncertainty $\delta_{f_N}$,
d) overall neutrino cross section uncertainty $\delta_{\sigma}$.
We consider following  values of the above systematic uncertainties:
 $\delta_{E} = 5\%$ with ${E}_0 = 1$ GeV for $E< 1$ GeV, and
 $\delta_{E} = 5\%$ with ${E}_0 = 10$ GeV for $E>10$ GeV,
 $\delta_z=4\%$ with $\cos\theta_0^{\rm zenith} =0.5$, 
  $\delta_{f_N}=10\%$, 
 and $\delta_{\sigma}=15\%$.
These are derived from the latest calculation of atmospheric neutrino flux \cite{Sanuki:2006yd}.
For each set of oscillation parameters we minimize the $\chi^2$ with 
respect to the all pull variables and then use these values to calculate the 
$\chi^2$ for that set of parameters.

We consider all uncertainties as a function of reconstructed
neutrino energy and zenith angle. Here we assumed that the tilt uncertainties
will not be changed too much due to the reconstruction. However, on the other
hand, if any tilt uncertainty
arises in reconstructed neutrino events from the reconstruction method, 
it is then accommodated in $\chi^2$.

To generate the theoretical data for chi-square analysis,
we first generate 500 years un-oscillated data for 1 Mton detector.
From this data we find the energy-angle correlated resolutions separately for $\nu_\mu$
and $\bar\nu_\mu$ (see Fig. 3 of \cite{Samanta:2008ag})
in 35 $E_\nu$ bins (in log scale for the range of $0.8-40$ GeV)
and 17 $\cos\theta^{\rm zenith}_\nu$ bins ($-1$ to $+1$). For a given grid
of $(E_\nu^i-\cos\theta_{\rm zenith}^j)$,
we calculate the efficiency of having $E_\mu \ge$ 0.8 GeV (threshold
of the detector).
For each set of oscillation parameters, we integrate the oscillated
atmospheric neutrino
flux folding the cross section, exposure time, target mass, efficiency
and resolution function to obtain the predicted data in the reconstructed
$\log E-L^{0.4}$ grid
\footnote{One can do this directly generating 500 Mton.year data (to ensure
the statistical error negligible) for {\bf each set} of oscillation
parameters and then calculating 1Mton.year equivalent data from it, which would
be the more straight forward method.
The marginalization study with this method is almost
an un-doable job in a  normal CPU. However, an exactly equivalent result is
obtained here using the energy-angle correlated resolution functions.}.
We use the Charge Current (CC) cross section of Nuance-v3 \cite{Casper:2002sd} and the
Honda flux  of 3-dimensional scheme \cite{Sanuki:2006yd}.
This method of obtaining the theoretical data in bins of muon energy and zenith angle 
has been used and discussed in our previous works \cite{Samanta:2008af,Samanta:2008ag,
Samanta:2006sj,Samanta:2009qw}.

We have done this study considering only the muons produced in the CC interactions. 
It is highly expected that the addition of hadron energy to the muon energy of an event will
improve the result. However, since the hadron energy resolution depends significantly on
the thickness of the iron plates, this addition of hadron energy will be more reliable in case of 
GEANT-based studies with realistic backgrounds.   

From GEANT simulation of ICAL detector
it is found that the energy resolution varies 4-10\% and angular resolution 4-12\% 
depending on the energy and the direction for our considered range of energy.
The width of these resolutions are very negligible compared to that obtained from 
kinematics of the scattering processes.
We have also checked from GEANT simulation that the wrong charge identification possibility of 
ICAL detector is also almost zero when the magnetic field is $\gapp 1$ Tesla for our 
considered range of muon energy. 

The iron plates are stacked horizontally and the muons produced in the horizontal direction
or very near to it cannot be detected. So we put a selection criteria in our analysis. The
muons for a given energy must be with a zenith angle such that
$|90^\circ-\theta_{\rm zenith}|$ is greater than   the half width at half maxima of the
scattering angle distribution with that energy. This is discussed in detail in \cite{Samanta:2008af}.

\begin{figure*}[htb]
\includegraphics[height=6cm,angle=0]{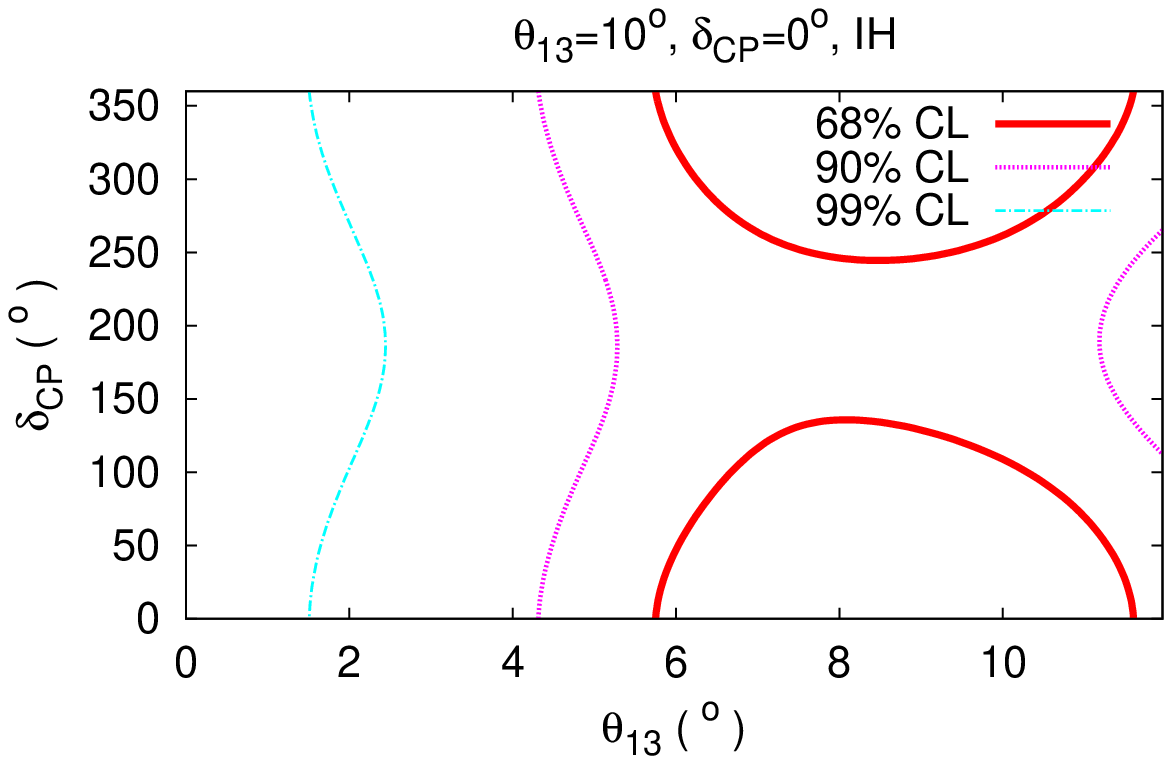}
\includegraphics[height=6cm,angle=0]{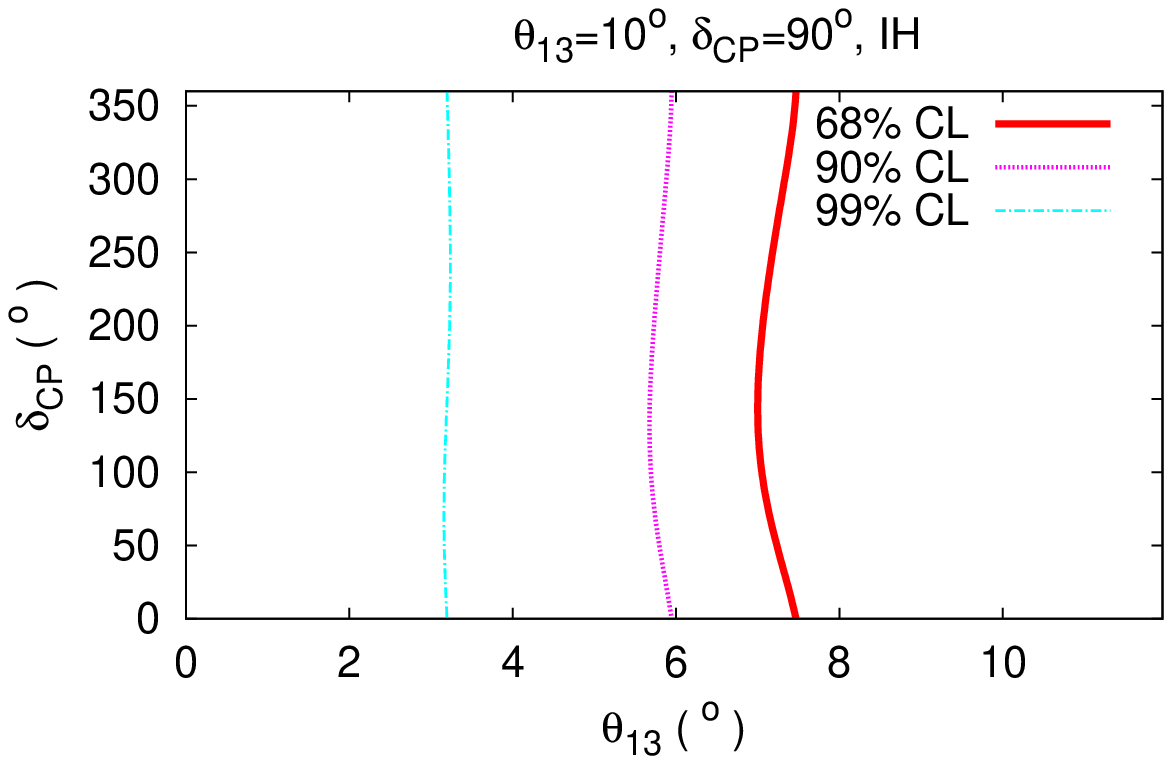}
\includegraphics[height=6cm,angle=0]{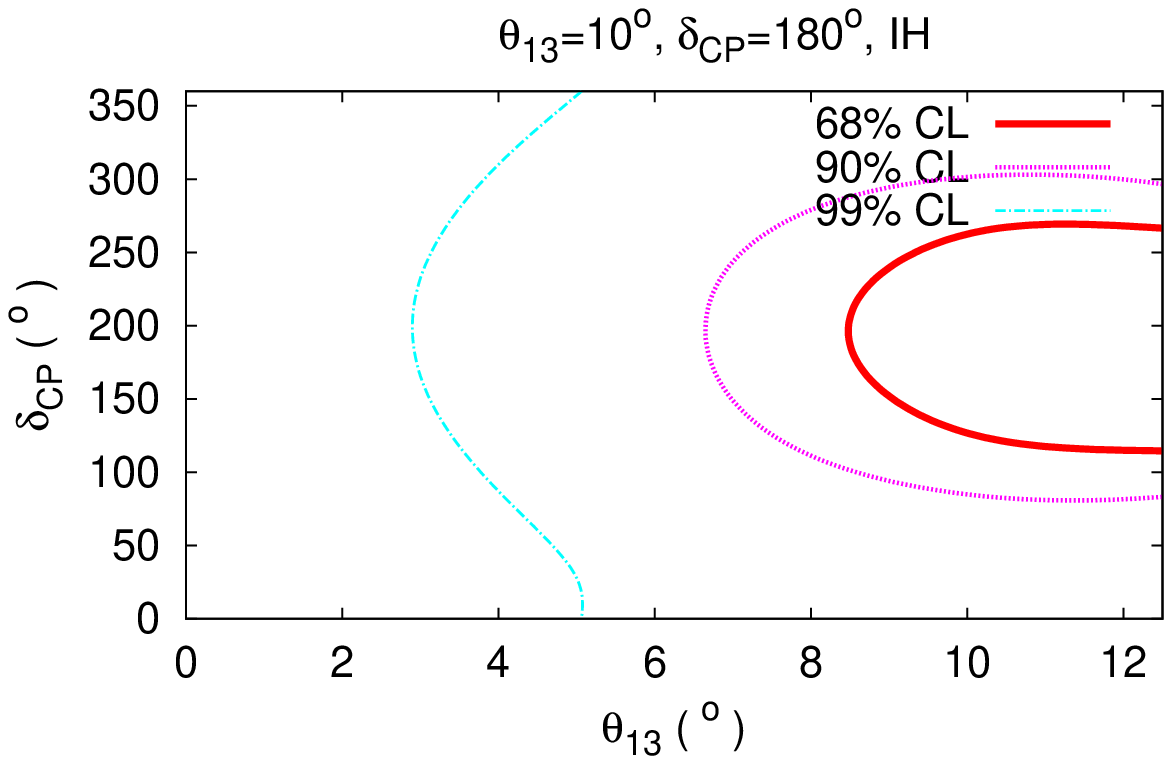}
\caption{\sf \small The  contours in $\delta_{CP}-\theta_{13}$ plane for the
inputs of IH and $\theta_{13}=10^\circ$ with $\delta_{CP}=0^\circ,~90^\circ$ and $180^\circ$,
respectively. 
\label{f:mc}
}
\end{figure*}
\section{Results}
\begin{figure*}[htb]
\includegraphics[height=6cm,angle=0]{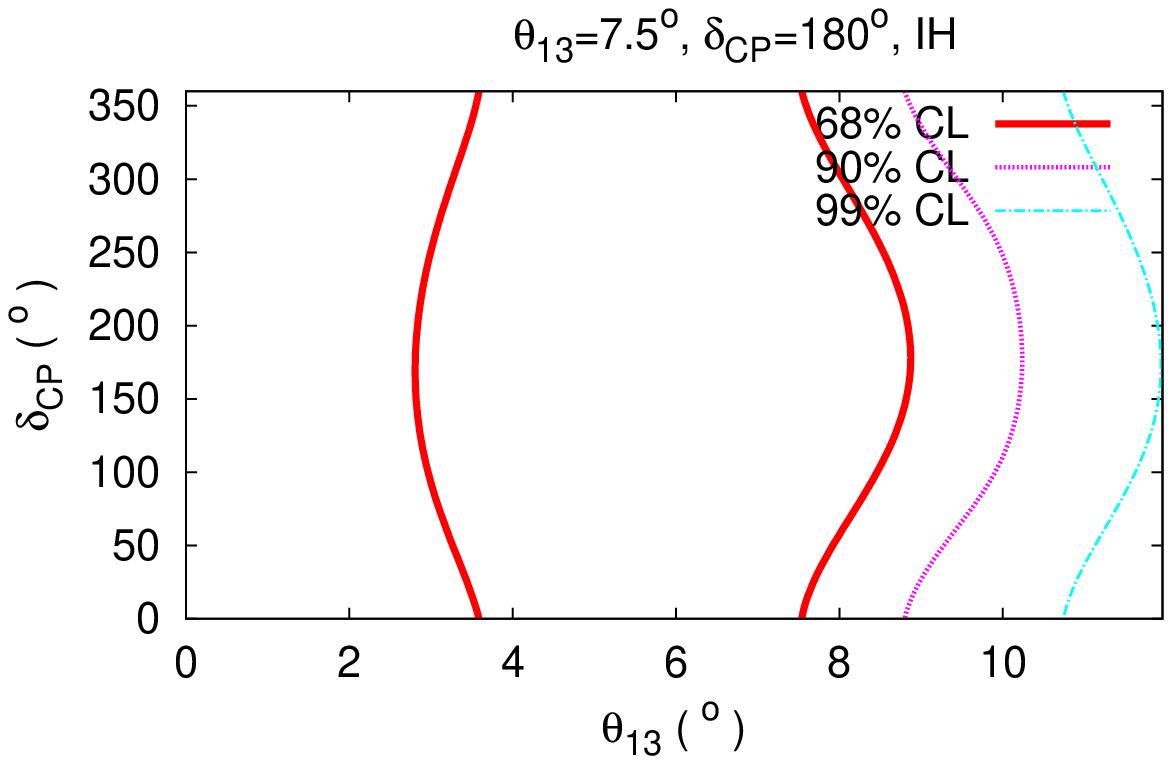}
\caption{\sf \small The same as Fig. \ref{f:mc}, but for different inputs
as shown in plot.
}
\label{f:mt7.5}
\end{figure*}

\begin{figure*}[htb]
\includegraphics[height=6cm,angle=0]{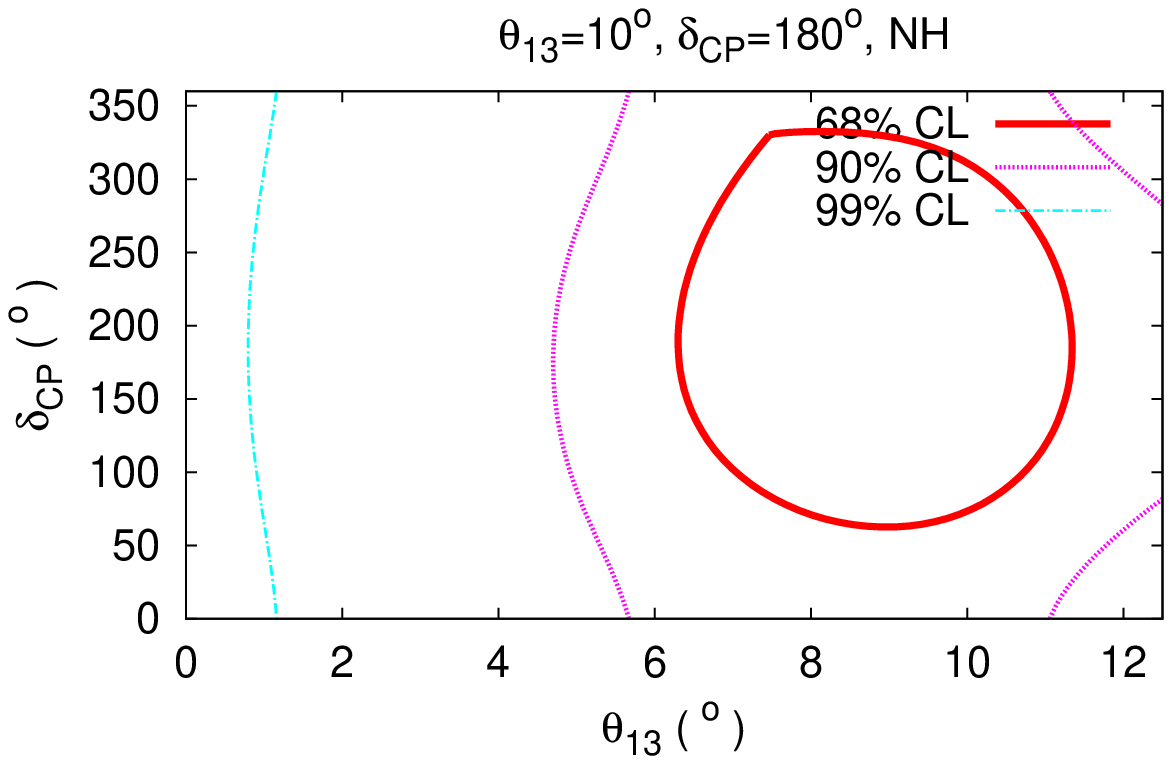}
\caption{\sf \small The same as Fig. \ref{f:mc}, but for different inputs
as shown in plot.
}
\label{f:pc180}
\end{figure*}

We have marginalized the $\chi^2=\chi^2_\mu+\chi^2_{\bar\mu}$ over all oscillation parameters
$\Delta m_{32}^2,~\theta_{23}, ~\theta_{13}$, $\delta_{\rm CP}$, $\Delta m_{21}^2$, and $\theta_{12}$
choosing  the range of $\Delta m_{32}^2=2.0-3.0\times 10^{-3}$eV$^2$,
$\theta_{23}=37^\circ - 54^\circ$, $\theta_{13}=0^\circ - 12.5^\circ$
and $\delta_{\rm CP}=0^\circ - 360^\circ$. 
We set the range of  $\Delta m_{21}^2=7.06 -8.34 \times 10^{-5}$eV$^2$ and 
$\theta_{12}=30^\circ - 40^\circ$. However, the effect of $\Delta m_{21}^2$ comes in 
subleading order in the oscillation probability when $E\sim$ GeV.
The 2-dimensional 68\%, 90\%, 99\%
confidence level (CL) allowed parameter spaces (APSs) are obtained by considering
$\chi^2=\chi^2_{\rm min}+2.48,~4.83,~9.43$, respectively. 
We show the contours in $\theta_{13} - \delta_{CP}$ plane in Fig. \ref{f:mc} for the inputs of IH
and $\theta_{13}=10^\circ$ with $\delta_{CP}=0^\circ, 90^\circ$ and $180^\circ$, respectively.
We see that both upper and lower bounds can be obtained for $\delta_{CP}=180^\circ$ at 90\% CL.
In Fig. \ref{f:mt7.5}, we show  the contours for  $\theta_{13}=7.5^\circ$ with $\delta_{CP}=180^\circ$.
The contours with inputs of NH  and $\delta_{CP}=180^\circ$ are also shown  in Fig. \ref{f:pc180}. 
Here, we find that both upper and lower bounds can be obtained with atmospheric neutrinos only 
when $\delta_{CP}\sim 0^\circ$ and $180^\circ$. 
However, we find no bounds if $\delta_{CP}=90^\circ$.

\begin{figure*}[htb]
\includegraphics[height=6cm,angle=0]{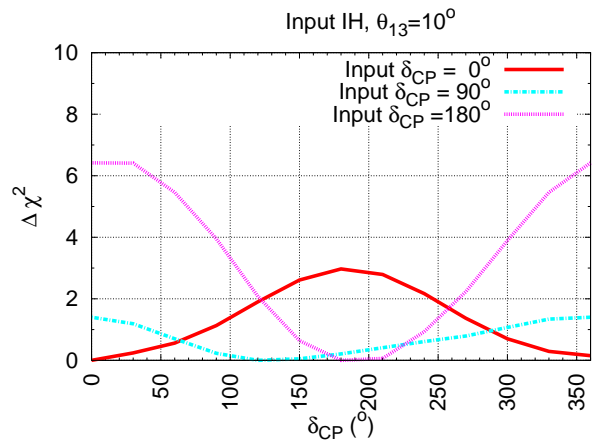}
\caption{\sf \small The variation of $\Delta \chi^2$ $[\chi^2 - \chi^2_{\rm min}]$ with $\delta_{CP}$ for different inputs of $\delta_{CP}$. 
The $\chi^2$ is marginalized over all oscillation parameters except $\delta_{CP}$. 
}
\label{f:absol}
\end{figure*}

\begin{figure*}[htb]
\includegraphics[height=6cm,angle=0]{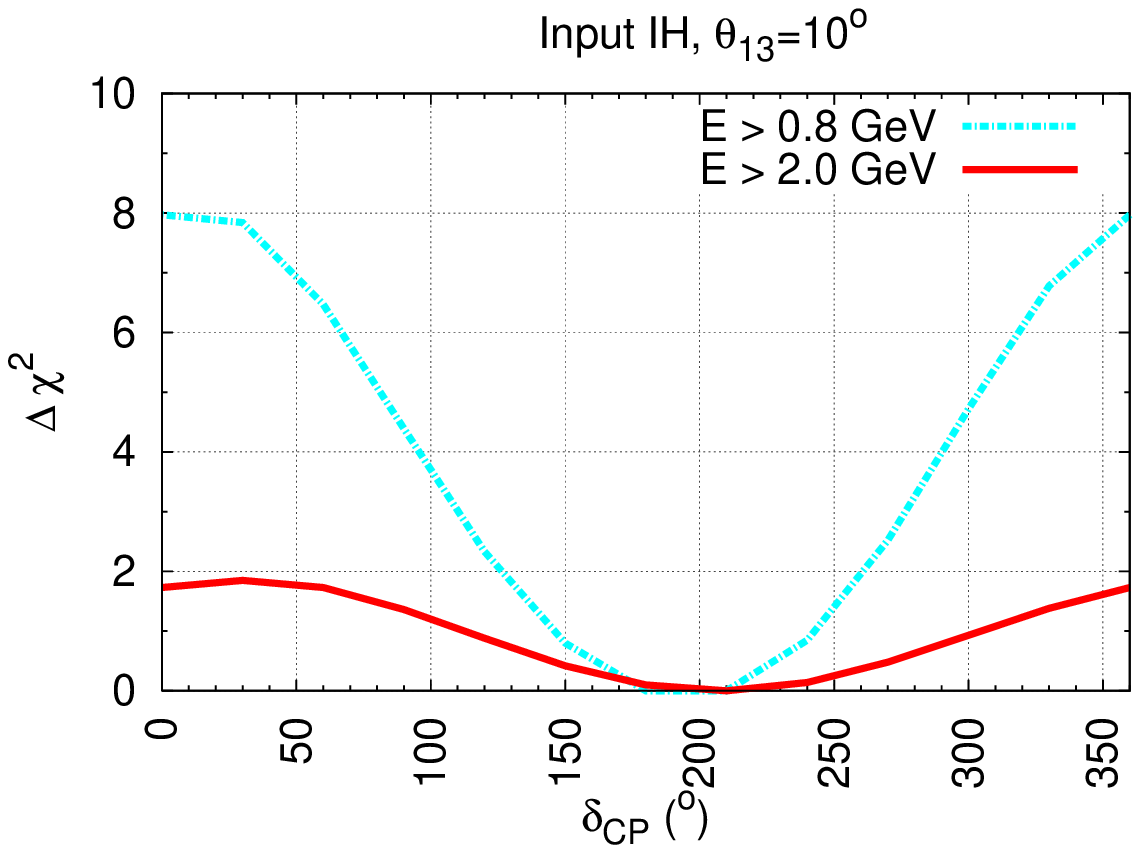}
\caption{\sf \small The variation of $\Delta \chi^2$ $[=\chi^2 - \chi^2_{\rm min}]$    with $\delta_{CP}$ keeping other oscillation parameters
fixed. We use linear binning in $E$ and $\cos\theta_{\rm zenith}$ with bin size  of 1 GeV and $0.066$,
respectively
for the chi-square analysis of this plot.
This binning is similar with that in \cite{Gandhi:2007td}. The lower solid curve represents the result with threshold of 2 GeV
used in \cite{Gandhi:2007td} and dashed one with threshold of 0.8 GeV used in our analysis.
  }
\label{f:threshold}
\end{figure*}

The absolute bounds on $\delta_{CP}$ for its different inputs are shown in Fig. \ref{f:absol}. 
This is obtained after marginalization over all oscillation parameters except $\delta_{CP}$. 
In Fig. \ref{f:threshold}, we see that the sensitivity falls drastically 
if one changes the threshold of the detector from 0.8 GeV to 2 GeV. This result
can be understood from Figs. \ref{f:nh} and \ref{f:ih}. 
Now we can conclude that the events at energy range $E\approx 1-2$ GeV contribute
mainly  in CP phase sensitivity in atmospheric neutrino oscillation. 

It should be notable here that the binning of the data to obtain the results in Fig. 
\ref{f:threshold}  is different from all other analyses in this paper. Here, we have used
linear binning in $E$ and $\cos\theta_{\rm zenith}$ with bin size of 1 GeV 
and $0.066$, respectively. However, we have enlarged the bin size of $E$ by adding the nearest bin in $E$
if number of events in a bin is $ \le 4$. This type of binning has been considered to compare our results with
that obtained in \cite{Gandhi:2007td}. If we use the same threshold of 2 GeV as used for the analysis 
in \cite{Gandhi:2007td}, the results are very similar in nature with \cite{Gandhi:2007td}. 
However, here  we have considered the energy range 2-35 GeV, while the range 2-8 GeV is used for analysis in  
\cite{Gandhi:2007td}.
From this study, one extra point we gain is that results does not change significantly for different
 types of binning. 
However, we have chosen the binning in $\log E - L^{0.4}$ plane for all other analyses. 
This was optimized for the precision study of $\Delta m^2_{32}$ and $\theta_{23}$
\cite{Samanta:2008ag}. 
But, here we checked that the type of binning does not change the results significantly.

\section{The case with $\delta_{CP}\approx 90^\circ$ or $270^\circ$}
At $\delta_{CP}= 90^\circ$ or $270^\circ$, there is practically no $\delta_{CP}$ dependence in 
the muon neutrino survival probability $P (\nu_\mu \rightarrow \nu_\mu)$ in 
symmetric matter profile \cite{Kimura:2002wd}. This can be understood in the following way.
The oscillation probabilities can be expressed as  linear combinations of 
cosine and sine functions of $\delta_{CP}$ \cite{Kimura:2002wd, Yokomakura:2002av}. 
The probability $P (\nu_\mu \rightarrow \nu_\mu)$ in symmetric matter profile can be expressed as 
\begin{equation}
P (\nu_\mu \rightarrow \nu_\mu) = A \cos\delta_{CP}  + D \cos 2\delta_{CP} + C,
\nonumber
\end{equation}
where, the coefficient $A, ~B$ and $C$ are functions of oscillation parameters, 
but independent of $\delta_{CP}$. The magnitude of $D$ is $O (\Delta m_{21}^2 \sin^2\theta_{13})$ 
and hence it is difficult to observe the effect of the term proportional to $\cos 2\delta_{CP}$.
So, effectively, $P (\nu_\mu \rightarrow \nu_\mu)$ is proportional to $\cos\delta_{CP}$ and 
there is no $\delta_{CP}$ dependence when $\delta_{CP}= 90^\circ$ or $270^\circ$.
The dependency grows very slowly as one goes away from these particular values and hence it is very 
difficult to measure $\delta_{CP}$ if its true value is around $90^\circ$ or $270^\circ$.  
However, due to slow increase of its dependence, one can expect the possibility
of its measurement at these values. In Fig. \ref{f:mc}, we have shown the case for 
$\delta_{CP}= 90^\circ$. We see there is almost no sensitivity.

We have studied further the possibilities of improving the sensitivity in different optimistic ways.
First, we try to see if any improvement of the systematic uncertainties can make it
possible to observe the CP violation. For this purpose
we have changed the uncertainties and  see the impact of them on APSs.
We see from the discussion in section \ref{s:flux} that the flux is less known 
for $E\le 1$ GeV. We have assumed a tilt uncertainty $\delta_E= 5\%$ with energy for this 
region. 
We have checked by improving the energy tilt uncertainty from 5\% to 2\% that
there is no much significant change in CP sensitivity, except a little hump around $90^\circ$. 
To check the effect of horizontal/vertical flux uncertainty,
we have changed the value of $\delta_z$ from 4\% to 2\%. Here also, we  find no such significant change.

If we decrease the threshold from 0.8 GeV to 0.6 GeV,
both CP sensitivity in oscillation probability (see Fig. \ref{f:nh} and \ref{f:ih}) and flux increases.
But, we have checked that there is also no  improvement in APSs with decreasing the threshold.
The fact is that with decrease in energy the angular resolution worsens very rapidly and
this kills the above prospects. 

At $\delta_{CP}\approx 90^\circ$ or $270^\circ$, we see from the discussion at the beginning
of this section that the CP sensitivity vanishes over whole $L-E$ space. It grows gradually
as one goes away from these values.
In Fig. \ref{f:nh} and \ref{f:ih} this feature is reflected and 
the difference $\Delta P$  appears relatively in the lower energy 
zone compared with $\delta_{CP}=180^\circ$ case.
In case of atmospheric neutrinos the direction is not fixed. 
The scattering angle between the muon and the neutrino is very large at this low energy and this
mainly acts against the CP phase measurements for $E < 1$ GeV. 
In case of the experiments with neutrino beams, the direction is known.
This gives the main advantage to measure the CP phase there. 
 
\section{Conclusion}
In this paper we have studied the sensitivity of a magnetized ICAL detector  in measuring
the CP phase with atmospheric neutrino oscillation. We have presented the results for 1 Mton.year
exposure of ICAL, which is 10 years running of the proposed 100 kTon detector.
Here, we have considered the muons (directly measurable) of the events, which are produced by the charge
current interactions generated by neutrino event generator Nuance-v3. We performed 
a marginalized chi-square study considering all possible systematic uncertainties.
We  find that one can be able to distinguish $\delta_{CP}\approx 0^\circ$ 
and $180^\circ$  at 90\% confidence level. However, there is no significant sensitivity for 
$\delta_{CP}\approx 90^\circ$ or $\delta_{CP}\approx 270^\circ$.

{\it Acknowledgments:}
This research has been supported by the fund from Neutrino Physics projects
at HRI.
The use of excellent cluster computational
facility installed from the fund of this project  is also gratefully acknowledged. 
The author is also grateful to Shahid Ali Farooqui for the help in installation of this 
new cluster. The work was initially carried out at the HRI main cluster facility.


\begin{thebibliography}{999}

\bibitem{Fukuda:1998mi}
  Y.~Fukuda {\it et al.}  [Super-Kamiokande Collaboration],
  Phys.\ Rev.\ Lett.\  {\bf 81}, 1562 (1998)
  [arXiv:hep-ex/9807003].

\bibitem{Haines:1986yf}
  T.~J.~Haines {\it et al.},
  Phys.\ Rev.\ Lett.\  {\bf 57}, 1986 (1986).



\bibitem{Hirata:1988uy}
  K.~S.~Hirata {\it et al.}  [KAMIOKANDE-II Collaboration],
  Phys.\ Lett.\  B {\bf 205}, 416 (1988).


\bibitem{Fogli:2008ig}
  G.~L.~Fogli {\it et al.},
  Phys.\ Rev.\  D {\bf 78}, 033010 (2008)
  [arXiv:0805.2517 [hep-ph]].

\bibitem{mixing}
Z. Maki, M. Nakagawa and S. Sakata, Prog. Theor. Phys. {\bf 28}, 870
(1962).
B. Pontecorvo, JETP {\bf 6} (1958) 429;
{\bf 7} (1958) 172;



\bibitem{Ardellier:2004ui}
  F.~Ardellier {\it et al.},
  arXiv:hep-ex/0405032.

\bibitem{Guo:2007ug}
  X.~Guo {\it et al.}  [Daya-Bay Collaboration],
  arXiv:hep-ex/0701029.

\bibitem{Kim:2008zzb}
  S.~B.~Kim  [RENO Collaboration],
  AIP Conf.\ Proc.\  {\bf 981}, 205 (2008)
  [J.\ Phys.\ Conf.\ Ser.\  {\bf 120}, 052025 (2008)].

\bibitem{Itow:2001ee}
  Y.~Itow {\it et al.}  [The T2K Collaboration],
  arXiv:hep-ex/0106019.

\bibitem{Ayres:2004js}
  D.~S.~Ayres {\it et al.}  [NOvA Collaboration],
  arXiv:hep-ex/0503053.

\bibitem{Huber:2009cw}
  P.~Huber, M.~Lindner, T.~Schwetz and W.~Winter,
  arXiv:0907.1896 [hep-ph].

\bibitem{Gandhi:2007td}
  R.~Gandhi, P.~Ghoshal, S.~Goswami, P.~Mehta, S.~U.~Sankar and S.~Shalgar,
  Phys.\ Rev.\  D {\bf 76}, 073012 (2007)
  [arXiv:0707.1723 [hep-ph]].


\bibitem{Casper:2002sd}
  D.~Casper,
  Nucl.\ Phys.\ Proc.\ Suppl.\  {\bf 112}, 161 (2002)
  [arXiv:hep-ph/0208030].

\bibitem{Athar:2006yb}
  M.~S.~Athar {\it et al.}  [INO Collaboration],



\bibitem{Alcaraz:2000vp}
  J.~Alcaraz {\it et al.}  [AMS Collaboration],
  Phys.\ Lett.\  B {\bf 490}, 27 (2000).

\bibitem{Alcaraz:2000zz}
  J.~Alcaraz {\it et al.}  [AMS Collaboration],
  Phys.\ Lett.\  B {\bf 494}, 193 (2000).
\bibitem{Sanuki:2000wh}
  T.~Sanuki {\it et al.},
  Astrophys.\ J.\  {\bf 545}, 1135 (2000)
  [arXiv:astro-ph/0002481].
\bibitem{Boezio:2002ha}
  M.~Boezio {\it et al.},
  Astropart.\ Phys.\  {\bf 19}, 583 (2003)
  [arXiv:astro-ph/0212253].

\bibitem{Sanuki:2006yd}
  T.~Sanuki, M.~Honda, T.~Kajita, K.~Kasahara and S.~Midorikawa,
  Phys.\ Rev.\  D {\bf 75}, 043005 (2007)
  [arXiv:astro-ph/0611201].





%


%
%
\bibitem{Samanta:2008af}
  A.~Samanta,
  arXiv:0812.4639 [hep-ph].

\bibitem{Samanta:2008ag}
  A.~Samanta,
  Phys.\ Rev.\  D {\bf 79}, 053011 (2009)
  [arXiv:0812.4640 [hep-ph]].

\bibitem{Samanta:2006sj}
  A.~Samanta,
  Phys.\ Lett.\  B {\bf 673}, 37 (2009)
  [arXiv:hep-ph/0610196].


\bibitem{Samanta:2009qw}
  A.~Samanta,
  arXiv:0907.3540 [hep-ph].

\bibitem{Kimura:2002wd}
  K.~Kimura, A.~Takamura and H.~Yokomakura,
  Phys.\ Rev.\  D {\bf 66}, 073005 (2002)
  [arXiv:hep-ph/0205295].


\bibitem{Yokomakura:2002av}
  H.~Yokomakura, K.~Kimura and A.~Takamura,
  Phys.\ Lett.\  B {\bf 544}, 286 (2002)
  [arXiv:hep-ph/0207174].



\end{thebibliography}
\end{document}